\documentclass[runningheads]{llncs}
\usepackage{graphicx}

\usepackage{amsmath}
\usepackage{amssymb}
\usepackage{mathrsfs} 
\usepackage{mathtools}  %to define floor and ceil bellow
\usepackage{dsfont} % for bold 1
\allowdisplaybreaks

\newcommand{\Exx}[2]{\mathbb{E}_{#1}\left(#2\right)}
\renewcommand{\Pr}[1]{\mathbb{P}\left(#1\right)}

\newcommand{\salg}[1]{\mathfrak {#1}}

\newcommand{\cf}{\mathds{1}}

\newcommand{\diff}[2]{\frac{\mathrm{d}  #1}{\mathrm{d}  #2}}

\newcommand{\norm}[1]{\left\|#1\right\|}

\newcommand{\absval}[1]{\left|#1\right|}

\def \d{\mbox{\(\,\mathrm{d}\)}}
 
\renewcommand{\epsilon}{\varepsilon}

\newcommand{\Rr}{\mathbb{R}}

\newcommand{\Nn}{\mathbb{N}}

\newcommand{\set}[2]{\{\,#1 \, : \, #2\,\} }
\newcommand{\bigset}[2]{\left\{\,#1 \, : \, #2\,\right\} }

\newcommand{\ent}[1]{H_{#1}}
\newcommand{\e}{\mathrm{e}}
\newcommand{\diam}{\operatorname{diam}}
\newcommand{\keyt}[1]{\emph{#1}}
\newcommand{\V}[1]{\mathbf{#1}}

\begin{document}
\title{On the entropy of rectifiable and stratified measures}  %\thanks{Supported by organization x.}}
\author{
Juan Pablo Vigneaux%\orcidID{1111-2222-3333-4444} }
}

\authorrunning{ J. P. Vigneaux}
\institute{
Department of Mathematics, California Institute of Technology, \\ Pasadena CA 91125, USA\\
\email{vigneaux@caltech.edu}}
\maketitle              % typeset the header of the contribution
\begin{abstract}
We summarize some results of geometric measure theory concerning rectifiable sets and measures. Combined  with the entropic chain rule for disintegrations  (Vigneaux, 2021), they   account for some  properties of the entropy of rectifiable measures with respect to the Hausdorff measure first studied by (Koliander et al., 
 2016). Then we present  some recent work on \emph{stratified measures}, which are convex combinations of rectifiable measures. These generalize discrete-continuous mixtures and may have a singular continuous part. Their entropy obeys a chain rule, whose ``conditional term''  is an average of the entropies of the rectifiable measures involved. We state an asymptotic equipartition property (AEP) for stratified measures that shows concentration on strata of a few ``typical dimensions'' and that links the conditional term  of the chain rule to the volume growth of typical sequences in each stratum. 
 %Finally, we discuss the relation with Renyi's information dimension and dimensional entropy. 
%We introduce a category of probabilistic finite categories and define two functions that share the characterizing properties of Shannon entropy. 

\end{abstract}
\section{Introduction}

The starting point of our considerations is the asymptotic equipartition property: 

\begin{proposition}[AEP]\label{prop:AEP}
Let $(E_X,\salg B,\mu)$ be a $\sigma$-finite measure space, and $\rho$ a probability measure on $(E_X,\salg B)$ such that  $\rho \ll \mu$. Suppose that the \emph{entropy}
\begin{equation}
    H_{\mu}(\rho) := -\int_{E} \ln \diff{\rho}{\mu} \d\rho =\Exx{\rho}{-\ln \diff{\rho}{\mu}}
\end{equation}
is finite. For every $\delta >0$, define the set of weakly $\delta$-typical realizations
\begin{equation}\label{eq:generalized_entropy}
W^{(n)}_\delta(\rho;\mu) =\bigset{(x_1,...,x_n)\in E_X^n}{\absval{-\frac 1n\ln \prod_{i=1}^n \diff\rho\mu(x_i) - \ent \mu(\rho)} < \delta}.
\end{equation}
Then,  for every $\epsilon>0$, there exists $n_0\in \Nn$ such that, for all $n\geq n_0$,
\begin{enumerate}
    \item $\Pr{W_\delta^{(n)}(\rho;\mu)}>1-\epsilon$ and
    \item $(1-\epsilon)  \exp\{n(\ent\mu(\rho)-\delta)\} \leq \mu^{\otimes n}(W_\delta^{(n)}(\rho;\mu)) \leq \exp\{n(\ent\mu(\rho)+\delta)\}. $
\end{enumerate}
\end{proposition}

The proof of this proposition only depends on the weak law of large numbers, which ensures the convergence in probability of $-\frac 1n \sum_{i=1}^n \ln f(x_i)$ to its mean $H_{\mu}(\rho)$, see \cite[Ch.~12]{Vigneaux2019-thesis} and \cite[Ch.~8]{Cover2006}.

Shannon's discrete entropy and differential entropy are particular cases of \eqref{eq:generalized_entropy}: the former when $\mu$ is the counting measure on a discrete set equipped with the $\sigma$-algebra of all subsets; the latter when $E_X=\Rr^d$, $\salg B$ is generated by the open sets, and $\mu$ is the Lebesgue measure. Up to a sign, the Kullback-Leibler divergence, also called \emph{relative entropy}, is another particular case, that arises when $\mu$ is a probability measure. 

One can imagine other examples,  geometric in nature, that involve measures $\mu$ on $\Rr^d$ that are singular continuous. For instance, $E_X$ could be a Riemannian manifold equipped with the Borel measure given by integration of its Riemannian volume form. The measure-theoretic nature of Proposition \ref{prop:AEP} makes the smoothness in this example  irrelevant. It is more natural to work with geometric measure theory.

\section{Some elements of Geometric Measure Theory}

Geometric measure theory ``could be described as differential geometry, generalized through measure theory to deal with maps and surfaces that are not necessarily smooth, and applied to the calculus of variations'' \cite{Morgan2008}.  The place of smooth maps is taken by Lipschitz maps, which are differentiable almost everywhere \cite[p. 46]{Ambrosio2000}. In turn, manifolds are replaced by rectifiable sets, and  the natural notion of volume for such sets is the Hausdorff measure. 

\subsection{Hausdorff measure and dimension}
To define the Hausdorff measure, recall first that the diameter of a subset $S$ of the Euclidean space $(\Rr^d,\norm{\cdot}_2)$ is  
$\diam(S) = \sup\set{\norm{x-y}_2}{x,y\in S}.$
 For any $m\geq 0$ and any  $A\subset \Rr^n$, define 
\begin{equation}\label{eq:Hausdorff_meas}
\mathcal H^m(A) = \lim_{\delta\to 0} \inf_{\{S_i\}_{i\in I}} \sum_{i\in I} w_m \left(\frac {\diam(S_i)}{2}\right)^m,
\end{equation}
where $w_m=\pi^{m/2}/\Gamma(m/2+1)$ is the volume of the unit ball $B(0,1)\subset \Rr^m$,  and the infimum is taken over all countable coverings $\{S_i\}_{i\in I}$ of $A$ such that each set $S_i$ has diameter at most $\delta$.  This is an outer measure and its restriction to  the Borel $\sigma$-algebra is a measure i.e. a $\sigma$-additive $[0,\infty]$-valued set-function  \cite[Thm. 1.49]{Ambrosio2000}. Moreover, the measure $\mathcal H^d$ equals the standard Lebesgue measure $\mathcal L^d$ \cite[Thm. 2.53]{Ambrosio2000} and $\mathcal H^0$ is the counting measure. More generally, when $m$ is an integer between $0$ and $d$, $\mathcal H^m$ gives a natural notion of $m$-dimensional volume. The $1$- and $2$-dimensional volumes coincide, respectively, with the classical notions of length and area, see Examples \ref{ex:curves} and \ref{ex:surfaces} below.

For a given set $A\subset \Rr^d$, the number $\mathcal H^m(A)$ is either $0$ or $\infty$, except possibly for a single value of $m$, which is then called the Hausdorff dimension $\dim_H A$ of $A$. More precisely \cite[Def. 2.51]{Ambrosio2000},
\begin{equation}
    \dim_H A := \inf\set{k\in[0,\infty)}{\mathcal H^k (A)=0} = \sup\set{k\in[0,\infty)}{\mathcal H^k (A) = \infty}.
\end{equation}

Finally, we remark here that the Hausdorff measure interacts very naturally with Lipschitz maps.

\begin{lemma}[\protect{\cite[Prop.~2.49]{Ambrosio2000}}]\label{lem:inequality_hausdorff_Lipschitz_maps}
If $f:\Rr^d\to \Rr^{d'}$ is a Lipschitz function,\footnote{A function $f:X\to Y$ between metric spaces $(X, d_X)$ and $(Y,d_Y)$ is called Lipschitz if there exists $C>0$ such that
$$\forall x,x'\in X,\quad d_Y(f(x),f(x')) \leq C d_X(x,x').$$
The Lipschitz constant  of $f$, denoted  $\operatorname{Lip}(f)$, is the smallest $C$ that satisfies this condition.
} 
with Lipschitz constant $\operatorname{Lip}(f)$, then for all  $m\geq 0$ and every subset $E$ of $\Rr^d$, $\mathcal H^m(f(E)) \leq \operatorname{Lip}(f)^k \mathcal H^m(E)$.
\end{lemma}

\subsection{Rectifiable sets}

\begin{definition}[\protect{\cite[3.2.14]{Federer1969}}] A subset $S$ of $\mathcal \Rr^d$ is called \emph{$m$-rectifiable} (for $m\leq d$) if it is the image of a bounded subset of $\Rr^m$ under a Lipschitz map and \emph{countably $m$-rectifiable} if it is a countable union of $m$-rectifiable sets.  The subset $S$ is  
 \keyt{countably $(\mathcal H^m,m)$-rectifiable}  if there exist countable $m$-rectifiable set containing $\mathcal H^m$-almost all of $S$, this is, if there are bounded sets $A_k \subset \Rr^m$ and Lipschitz functions $f_k : A_k \to \Rr^d$, enumerated by $k\in \Nn$, such that $\mathcal H^m(S\setminus \bigcup_k f_k(A_k))=0$. 
 \end{definition}
 
 By convention, $\Rr^0$ is a point, so  that a countably $0$-rectifiable set is simply a countable set. An example of countable  $m$-rectifiable set is an $m$-dimensional $C^1$-submanifold $E$ of $\Rr^d$, see  \cite[App. A]{Alberti2019}. A set that differs from an $m$-dimensional $C^1$-submanifold by a  $\mathcal H^m$-null set is countably $(\mathcal H^m,m)$-rectifiable.

 For every countably $(\mathcal H^m,m)$-rectifiable set $E\subset \Rr^d$, there exists (see \cite[pp. 16-17]{Vigneaux-stratified} and the references therein) an $\mathcal H^m$-null set $E_0$, compact sets $(K_i)_{i\in \Nn}$ and injective Lipschitz functions $(f_i:K_i\to \Rr^d)_{i\in \Nn}$ such that the sets $f_i(K_i)$ are pairwise disjoint and
  \begin{equation}\label{eq:representation_rect_compact}
      E\subset E_0 \cup \bigcup_{i\in \Nn} f_i (K_i).
  \end{equation}

 It follows from Lemma \ref{lem:inequality_hausdorff_Lipschitz_maps} and the boundedness of the sets $K_i$ that
every $(\mathcal H^m,m)$-rectifiable set has $\sigma$-finite $\mathcal H^m$-measure. Because of monotonicity of $\mathcal H^m$,  any subset of an $(\mathcal H^m,m)$-rectifiable set is $(\mathcal H^m,m)$-rectifiable. Also, the countable union of $(\mathcal H^m,m)$-rectifiable subsets is $(\mathcal H^m,m)$-rectifiable.

The area and coarea formulas may be used to compute integrals on an $m$-rectifiable set with respect to the $\mathcal H^m$-measure. As a preliminary, we define the area and coarea factors.  Let $V$, $W$ be finite-dimensional Hilbert spaces and $L:V\to W$ be a linear map. Recall that the inner product gives explicit identifications $V\cong V^*$ and $W\cong W^*$ with their duals. 
\begin{enumerate}
    \item If $k:=\dim V \leq \dim V$, the \emph{$k$-dimensional Jacobian} or \emph{area factor} \cite[Def. 2.68]{Ambrosio2000} is $ J_k L = \sqrt{\det(L^* \circ L)}$,
    where $L^*:W^* \to V^*$ is the transpose of $L$.

    \item If $\dim V \geq \dim W=: d$, then the \emph{$d$-dimensional coarea factor} \cite[Def. 2.92]{Ambrosio2000} is $ C_d L =\sqrt{\det(L \circ L^*)}$.
\end{enumerate}
\begin{proposition}[Area formula, cf. \protect{\cite[Thm. 2.71]{Ambrosio2000}  and \cite[Thm. 3.2.3]{Federer1969}}]\label{Prop:area_formula}
Let $k,d$ be integers such that $k\leq d$ and $f:\Rr^k \to \Rr^d$ a Lipschitz function. For any Lebesgue measurable subset $E$ of $\Rr^k$ and $\mathcal L^k$-integrable function $u$, the function $y\mapsto \sum_{x\in E\cap f^{-1}(y)} u(x)$ on $\Rr^d$ is $\mathcal H^k$-measurable, and
\begin{equation}\label{eq:area_formula_functions}
\int_E u(x) J_kf(x) \d \mathcal L^k(x) = \int_{\Rr^d} \sum_{x\in E\cap f^{-1}(y)} u(x) \d \mathcal H^k(y).
\end{equation}
\end{proposition}

This implies in particular that, if $f$ is injective on $E$, then $f(E)$ is $\mathcal H^k$-measurable and $\mathcal H^k(f(E)) = \int_E  J_kf(x) \d \mathcal L^k(x)$. 

\begin{example}[Curves]\label{ex:curves}
Let $\varphi:[0,1]\to \Rr^d, \: t\mapsto (\varphi_1(t),...,\varphi_d(t))$ be a $C^1$-curve; recall that a $C^1$-map defined on a compact set is Lipschitz.  Then $d\varphi(t)= (\varphi_1'(t),...,\varphi_d'(t))$, and $J_1 \varphi = \sqrt{(d\varphi)^* (d\varphi)} = \norm{d\varphi}_2$. So we obtain the standard formula for the length of a curve, 
\begin{equation}
\mathcal H^1(\varphi([0,1])) = \int_0^1 \norm{d\varphi(t)}_2 \d t.
\end{equation}
\end{example}

\begin{example}[Surfaces]\label{ex:surfaces}
Let $\varphi = (\varphi_1, \varphi_2,\varphi_3):V\subset \Rr^2 \to \Rr^3$ be a Lipschitz, differentiable, and injective map defining a surface. In this case, $d\varphi(u,v) = \begin{pmatrix} \varphi_x(u,v) & \varphi_y(u,v)\end{pmatrix}$ where $\varphi_x(u,v)$ is the column vector $$(\partial_x\varphi_1(u,v), \partial_x\varphi_2(u,v),\partial_x\varphi_3(u,v))$$ and similarly for $\varphi_y(u,v)$. Then, if  $\theta$ denotes  the angle between $\varphi_x(u,v)$ and $\varphi_y(u,v)$,
\begin{equation}
J_2\varphi = \sqrt{\det\begin{pmatrix}
\norm{\varphi_x}^2 & \varphi_x \bullet \varphi_y \\
\varphi_x \bullet \varphi_y & \norm{\varphi_y}^2
\end{pmatrix}}= \norm{\varphi_x}\norm{\varphi_y} \sqrt{1-\cos\theta} = \norm{\varphi_x \times \varphi_y}.
\end{equation}
 Therefore,
\begin{equation}
\mathcal H^2(\varphi(V)) = \int_V \norm{\varphi_x \times \varphi_y(u,v)} \d u\d v,
\end{equation}
which is again the classical formula for the area of a parametric surface.
\end{example}

In many situations, it is useful to compute an integral over a countably $(\mathcal H^m,m)$-rectifiable set $E\subset \Rr^k$ as an iterated integral, first over the level sets $E\cap \set{x}{f(x)=t}$ of a Lipschitz function $f:\Rr^k\to \Rr^d$ with $d\leq k$, and then over $t\in \Rr^d$. In particular, if $E=\Rr^k$ and $f$ is a projection onto the space generated by some vectors of the canonical basis of $\Rr^k$, this procedure corresponds to Fubini's theorem. Its generalization to the rectifiable case is the coarea formula.

\begin{proposition}[Coarea formula, \protect{\cite[Thm. 2.93]{Ambrosio2000}}]
    Let $f:\Rr^k\to \Rr^d$ be a Lipschitz function, $E$  a countably $(\mathcal H^m,m)$-rectifiable subset of $\Rr^k$ (with $m\geq d$) and $g:\Rr^k\to [0,\infty]$ a Borel function. Then, the set $E\cap f^{-1}(t)$ is countably $(\mathcal H^{m-d},m-d)$-rectifiable and $\mathcal H^{m-d}$-measurable, the function $t\mapsto \int_{E\cap f^{-1}(t)} g(y) \d\mathcal H^{m-d}$ is $\mathcal L^d$-measurable on $\Rr^d$, and
    \begin{equation}\label{eq:coarea}
        \int_E g(x) C_d d^E f_x \d\mathcal H^m(x) = \int_{\Rr^d} \left(\int_{f^{-1}(t)} g(y) \d\mathcal H^{m-d}(y) \right) \d t.
    \end{equation}
\end{proposition}
Here $d^E f_x $ is the \emph{tangential differential} \cite[Def. 2.89]{Ambrosio2000}, the restriction of $df_x$ to the approximate tangent space to $E$ at $x$. The precise computation of this function is not essential here, but rather the fact that  $C_k d^E f_x >0$ $\mathcal H^m$-almost surely, hence $\mathcal H^{m}|_{ E}$ has a $(f,\mathcal L^d)$-disintegration given by the measures\\ $\{(C_k d^E f)^{-1} \mathcal H^{m-d}|_{ E \cap f^{-1}(t)}\}_{t\in \Rr^k}$, which are well-defined $\mathcal L^d$-almost surely.\footnote{Given a measure $\mu$ on a $\sigma$-algebra $\salg B$ and $B\in \salg B$, $\mu|_B$ denotes the restricted measure $A\mapsto \mu|_B(A) := \mu(A\cap B)$.}

\begin{remark}[On disintegrations]\label{rmk:disintegrations} Disintegrations are an even broader generalization of Fubini's theorem. 

Let $T:(E,\salg B)\to (E_T,\salg B_T)$ be a measurable map and let $\nu$ and $\xi$ be  $\sigma$-finite measures on $(E,\salg B)$ and $(E_T,\salg B_T)$ respectively. The measure $\nu$ has a $(T,\xi)$-disintegration $\{\nu_t\}_{t\in E_T}$ if
\begin{enumerate}
\item $\nu_t$ is a $\sigma$-finite measure on $\salg{B}$ such that $\nu_t(T\neq t)=0$ for $\xi$-almost every $t$;
\item for each measurable nonnegative function $f:E\to \Rr$, the map $t\mapsto \int_E f \d \nu_t$ is measurable, and $\int_E f\d \nu= \int_{E_T} \left(\int_E  f(x) \d\nu_t(x) \right)\d \xi(t)$.
\end{enumerate}
In case such a disintegration exists,  any probability measure $\rho = r \cdot \nu$ has a $(T,T_*\rho)$-disintegration $\{ \rho_t\}_{t\in E_t}$ such that each $\rho_t$ is a probability measure with density $r/\int_{E} r \d \nu_t $ w.r.t. $\nu_t$, and the following chain rule holds \cite[Prop. 3]{Vigneaux-GSI21-disintegrations}:
 \begin{equation}\label{eq:generalized_chain_rule}
 \ent{\nu}(\rho) =  \ent{\xi}(T_*\rho) +  \int_{E_T} \ent{\nu_t}(\rho_t)\d T_*\rho(t).
 \end{equation}
\end{remark}

\section{Rectifiable measures and their entropy}

Let $\rho$ be a locally finite measure and $s$ a nonnegative real number. 
Marstrand proved that if the limiting density $
\Theta_s(\rho,x) := 
\lim_{r\downarrow 0} \rho(B(x,r))/(w_s r^s)$
exists and is strictly positive and finite for $\rho$-almost every $x$, then $s$ is an integer not greater than $n$. Later Preiss proved that such a measure is also $s$-rectifiable in the sense of the following definition. For details, see e.g. \cite{DeLellis2008}.

\begin{definition}[\protect{\cite[Def.~16.6]{Mattila1995}}]\label{def:rectifiable_measure} A Radon outer measure $\nu$ on $\Rr^d$ is called $m$-rectifiable if $\nu \ll \mathcal H^m$ and there exists a countably $(\mathcal H^m,m)$-rectifiable Borel set $E$ such that $\nu(\Rr^d\setminus E)=0$.
\end{definition}

The study of these measures from the viewpoint of information theory, particularly the properties of the entropy $H_{\mathcal H^m}(\rho)$ of an $m$-rectifiable probability measure $\rho$, was carried out relatively recently by Koliander, Pichler, Riegler, and Hlawatsch in \cite{Koliander2016}. We provide here an idiosyncratic summary of some of their results. 

First, remark that in virtue of \eqref{eq:representation_rect_compact},  an $m$-rectifiable measure $\nu$ is absolutely continuous with respect to the restricted measure $\mathcal H^m |_{E^*}$, where $E^*$ is countably $m$-rectifiable and has the form $\bigcup_{i\in \Nn} f_i(K_i)$ with $f_i$ injective and $K_i$ Borel and bounded. %\footnote{By $\mu|_A$ we mean restriction of a measure $\mu:\salg B\to [0,\infty]$ to a set $A\in \salg B$, that is, $\mu|_A(B) = \mu(A\cap B)$ for all $B\in \salg B$. It is sometimes denoted $\mu\mres A$.}  We call a set $E^*$ of this form a \emph{carrier} (of $\nu$).  It is a set of Hausdorff dimension $m$ \cite[lem. 5]{Vigneaux-stratified}.
%Each set $f_i(K_i)$ is Borel, being the image of a Borel set under an injective map. 
 (A refinement of this construction gives a similar set such that, additionally, the density of $\rho$ is strictly positive \cite[App.~A]{Koliander2016}.) Although the product of an $(\mathcal H^{m_1},m_1)$-rectifiable set and an $(\mathcal H^{m_2},m_2)$ rectifiable set is \emph{not} $(\mathcal H^{m_1+m_2},m_1+m_2)$-rectifiable|see \cite[3.2.24]{Federer1969}|the carriers behave better.  

\begin{lemma}[See \protect{\cite[Lem.~27]{Koliander2016}} and \protect{\cite[Lem. 6]{Vigneaux-stratified}}]\label{lemma:products_carriers}
If $S_i$ is a carrier of an $m_i$-rectifiable measure $\nu_i$ (for $i=1,2$), then $S_1\times S_2$ is  a carrier of $\nu_1\otimes\nu_2$, of Hausdorff dimension $m_1+m_2$. 
Additionally, the Hausdorff measure $\mathcal H^{m_1+m_2}|_{S_1\times S_2}$ equals $\mathcal H^{m_1}|_{S_1} \otimes \mathcal H^{m_2}|_{S^2}$. 
\end{lemma}

Let $\rho$ be an $m$-rectifiable measure, with carrier $E$ (we drop the $*$ hereon). It holds that $\rho \ll\mathcal H^m|_{E}$ and $\mathcal H^m|_E$ is $\sigma$-finite. If moreover $H_{\mathcal H^m|_{E}}(\rho)<\infty$,  Proposition \ref{prop:AEP} gives estimates for  $(\mathcal H^{m}|_E)^{\otimes n}(\mathcal  W^{(n)}_\delta (\rho; \mathcal H^m|_{E}))$. Lemma \ref{lemma:products_carriers} tells us that $E^n$ is $mn$-rectifiable and that $(\mathcal H^{m}|_E)^{\otimes n} = \mathcal H^{mn}|_{E^n}$, which is desirable because the Hausdorff dimension of $E^n$ is $mn$ and $\mathcal H^{mn}$ is the only nontrivial measure on it as well as on $W^{(n)}$, which as a subset of $E^n$ is $mn$ rectifiable too.

To apply the AEP we need to compute $H_{\mathcal H^m|_{E}}(\rho)$. In some cases, one can use the area formula (Proposition \ref{Prop:area_formula}) to ``change variables'' and express $H_{\mathcal H^m|_{E}}(\cdot)$ in terms of the usual differential entropy. For instance, suppose $A$ is a bounded Borel subset of $\Rr^k$ of nontrival $\mathcal L^k$-measure and  $f$ is an injective Lipschitz function on $A$. The set   $f(A)$ is $k$-rectifiable. Moreover, if $\rho$ is a probability measure such that $\rho \ll \mathcal L^k|_{A}$ with density $r$, then the area formula  applied to $u=r/J_k f$ and $E=f^{-1} (B)$, for some Borel subset $B$ of $\Rr^d$, shows that $f_*\rho \ll \mathcal H^k|_{ f(A)}$ with density $(r/J_k f) \circ f^{-1}$, which is well-defined $(\mathcal H^k|_{ f(A)})$-almost surely. A simple computation yields
\begin{equation}
    H_{\mathcal H^k|_{f(A)}}(f_*\rho) = H_{\mathcal L^k}(\rho) + \Exx{\rho}{\ln J_k f}.
\end{equation}
There is a more general formula of this kind when $A$ is a rectifiable subset of $\Rr^d$.

Finally, we deduce the chain rule for the entropy of rectifiable measures as a consequence of our general theorem for disintegrations (Remark \ref{rmk:disintegrations}). Let $E$ be a countably  $(\mathcal H^m,m)$-rectifiable subset of $\Rr^k$, $f:\Rr^k\to \Rr^d$  a Borel function (with $d\leq m$), and $\rho$ a probability measure such that $\rho \ll \mathcal H^m|_E$. Because $\mathcal H^m |_E$ has an $(f,\mathcal L^d)$-disintegration $\{F^{-1} \mathcal H^{m-d}|_{E\cap f^{-1}(t)}\}_{t\in \Rr^k}$, with $F=C_d d^E f$, then $f_*\rho \ll \mathcal L^d$, and 
\begin{equation}\label{eq:chain_rule_coarea}
    H_{\mathcal H^m|_E} (\rho) = H_{\mathcal L^d}(f_*\rho) + \int_{\Rr^k} H_{F^{-1} \mathcal H^{m-d}|_ {E\cap f^{-1}(t))}}(\rho_t) \d f_*\rho. 
\end{equation}
The probabilities $\rho_t$ are described in Remark \ref{rmk:disintegrations}. If one insists in only using the Hausdorff measures as reference measures, one must rewrite the integrand in \eqref{eq:chain_rule_coarea} using the chain rule for the Radon-Nikodym derivative:
\begin{align*}
    H_{F^{-1} \mathcal H^{m-d} |_{E\cap f^{-1}(t)}}(\rho_t) &= \Exx{\rho_t}{-\ln \diff{\rho_t}{\mathcal H^{m-d}|_{E\cap f^{-1}(t)}} \diff{\mathcal H^{m-d}|_{E\cap f^{-1}(t)}}{F^{-1} \mathcal H^{m-d}|_{E\cap f^{-1}(t)}}}\\
    &= \Exx{\rho_t}{-\ln \diff{\rho_t}{\mathcal H^{m-d}|_{E\cap f^{-1}(t)}} } - \Exx{\rho_t}{\ln F}.
\end{align*}
One recovers in this way the formula (50) in \cite{Koliander2016}.

\section{Stratified measures}

\begin{definition}[$k$-stratified measure]\label{def:stratified_meas}
A measure $\nu$ on $(\Rr^d,\mathcal B(\Rr^d))$ is \emph{$k$-stratified}, for $k\in \Nn^*$,  if there are integers $(m_i)_{i=1}^k$ such that  $0 \leq m_1 < m_2 <... <m_k \leq d$ and $\nu$ can be expressed as a sum $\sum_{i=1}^k \nu_i$, where each $\nu_i$ is a nonzero $m_i$-rectifiable measure.
\end{definition}

Thus  $1$-stratified measures are rectifiable measures. If $\nu$ is $k$-stratified for some $k$ we simply say that $\nu$ is a \emph{stratified measure.}

A fundamental nontrival example to bear in mind is a discrete-continuous mixture, which corresponds to  $k=2$, $E_1$ countable, and $E_2 = \Rr^d$. More generally, a stratified measure has a Lebesgue decomposition with a singular continuous part provided some $m_i$ is strictly between $0$ and $d$.

Let $\rho$ be a probability measure that is stratified in the sense above. We can always put it in the \emph{standard form} 
$
\rho = \sum_{i=1}^k q_i \rho_i$, 
where each $\rho_i$ is a rectifiable probability measure with carrier $E_i$ of dimension $m_i$ (so that $\rho_i = \rho_i|_{ E_i}$), the carriers $(E_i)_{i=1}^k$ are disjoint, $0 \leq m_1 < \cdots < m_k \leq d$, and $(q_1,...,q_k)$ is a probability vector \emph{with strictly positive entries.} The carriers can be taken to be disjoint because if $E$ has Hausdorff dimension $m$, then $\mathcal H^k(E)=0$  for $k>m$, hence one can prove \cite[Sec. IV-B]{Vigneaux-stratified} that $E_i\setminus (\bigcup_{j=1}^{i-1} E_j)$ is a carrier for $\nu_i$, for $i=2,...,k$.

We can regard $\rho$ as the law of a random variable $X$ valued in $E_X:=\bigcup_{i=1}^k E_i$ and the vector $(q_1,...,q_k)$ as the law $\pi_*\rho$ of the discrete random variable $Y$ induced by the projection $\pi$ from  $E_X$ to $E_Y:=\{1,...,k\}$ that maps $x\in E_i$ to $i$. We denote by $D$ the random variable $\dim_H E_Y$, with expectation $\mathbb E(D)=\sum_{i=1}^k m_i q_i$.

The measure  $\rho$ is absolutely continuous with respect to $\mu = \sum_{i=1}^k \mu_i$, where $\mu_i = \mathcal H^{m_i}|_{ E_i}$, so it makes sense to consider the entropy $H_{\mu}(\rho)$; it has a concrete probabilistic meaning in the sense of Proposition \ref{prop:AEP}. Moreover, one can prove that $\diff{\rho}{\mu} = \sum_{i=1}^m q_i \diff{\rho_i}{\mu_i} \cf_{E_i}$ \cite[Lem. 3]{Vigneaux-stratified} and therefore 
\begin{equation}\label{eq:chain_rule}
\ent{\mu}(\rho) = \ent{}(q_1,...,q_n) + \sum_{i=1}^k q_i \ent{\mu_i}(\rho_i)
\end{equation}
holds \cite[Lem. 4]{Vigneaux-stratified}. This formula also follows form the chain rule for general disintegrations (Remark \ref{eq:chain_rule_coarea}), because $\{\rho_i\}_{i\in E_Y}$ is a $(\pi,\pi_*\rho)$-disintegration of $\rho$. 

 The powers of $\rho$ are also stratified. In fact, remark that 
 \begin{equation}
\rho^{\otimes n} = \sum_{\V y=(y_1,...,y_n) \in E_Y^n}  q_1^{N(1;\V y)} \cdots q_k^{N(k;\V y)} \rho_{y_1}\otimes \cdots \otimes \rho_{y_n}
\end{equation}
where $N(a;\V y)$ counts the appearances of the symbol $a\in E_Y$ in the word $\V y$. Each measure $\rho_{\V y} := \rho_{y_1}\otimes \cdots \otimes \rho_{y_n}$ is absolutely continuous with respect to $\mu_{\V y}:= \mu_{y_1} \otimes \cdots \otimes \mu_{y_n}$. It follows from Lemma \ref{lemma:products_carriers}  that for any $\V y\in E_Y^n$,  the \emph{stratum} $\Sigma_{\V y}:= E_{y_1}\times \cdots \times E_{y_n}$ is also a carrier, of dimension $m(\V y) := \sum_{j=1}^n \dim_H E_{y_j}$, and the product measure $\mu_{\V y}$ equals $\mathcal H^{m(\V y)}|_{\Sigma_{\V y}}$. Therefore each measure $\rho_{\V y}$ is rectifiable. We can group together the $\rho_\V y$ of the same dimension to put $\rho^{\otimes n}$ as in Definition \ref{def:stratified_meas}.

By Proposition \ref{prop:AEP}, one might approximate $\rho^{\otimes n}$ with an arbitrary level of accuracy by its restriction to the weakly typical realizations of $\rho$, provided $n$ is big enough. In order to get additional control on the dimensions appearing in this approximation, we restrict it further, retaining only the strata that correspond to \emph{strongly} typical realizations of the random variable $Y$. 

Let us denote by $Q$ the probability mass function (p.m.f) of $\pi_*\rho$. Recall that $\V y\in E_Y^n$ induces a probability law $\tau_{\V y}$ on $E_Y$, known as \emph{empirical distribution},  given by 
$\tau_{\V y}(\{a\}) = N(a;\V y)/n.$  Csisz\'ar and K\"orner \cite[Ch.~2]{Csiszar1981} define $\V y\in E_Y^n$ to be \emph{strongly $(Q,\eta)$-typical} if $\tau_{\V y}$, with p.m.f. $P$, is such that  $\tau_{\V y} \ll \pi_*\rho$ and, for all $a\in E_Y$, $|P(a)-Q(a)|<\eta$. We denote by $A_{\delta'_n}^{(n)}$ the set of these sequences when $\eta_n = n^{-1/2+\xi}$. In virtue of the union bound and Hoeffding's inequality, $(\pi_*\rho)^{\otimes n}(A_{\delta'_n}^{(n)})\geq 1-\epsilon_n$, 
where $\epsilon_n = 2|E_Y|\e^{-2n\eta_n^2}$; the choice of $\eta_n$ ensures that $\epsilon_n\to 0$ as $n\to\infty$. Moreover, the continuity of the discrete entropy in the total-variation distance implies that $A^{(n)}_{\delta'_n}$ is a subset of $W^{(n)}_{\delta'_n}$ with $\delta_n'=-|E_Y|\eta_n\ln \eta_n$, which explains our notation. See \cite[Sec. III-D]{Vigneaux-stratified}

We introduce the set  $T^{(n)}_{\delta,\delta'_n}=(\pi^{\times n})^{-1}(A_{\delta'_n}^{(n)})\cap W_{\delta}^{(n)}(\rho)$ of   \emph{doubly  typical sequences} in $E_X^n$, and call $T^{(n)}_{\delta,\delta'_n}(\V y) = T^{(n)}_{\delta,\delta'_n}\cap (\pi^{\times n})^{-1}(\V y)$ a \emph{doubly typical stratum} for any $\V y\in A_{\delta'_n}^{(n)}$. 

The main result of \cite{Vigneaux-stratified} is a refined version of the AEP for stratified measures that gives an  interpretation for the conditional term in the chain rule \eqref{eq:chain_rule}. 

\begin{theorem}
(Setting introduced above) For any $\epsilon > 0$ there exists an $n_0\in \Nn$ such that for any $n\geq n_0$ the restriction  of $\rho$ to $T^{(n)}_{\delta,\delta'_n}$, 
 $\rho^{(n)}=\sum_{ \V y \in A_{\delta'}^{(n)}} \rho^{\otimes n}|_{T^{(n)}_{\delta,\delta'}(\V y)},$ satisfies $d_{TV}(\rho^{\otimes n},\rho^{(n)})<\epsilon$.
  Moreover, the measure $\rho^{(n)}$ equals a sum of $m$-rectifiable measures for $m\in [n \mathbb E(D) - n^{1/2+\xi}, n\mathbb E(D) + n^{1/2+\xi}]$. The conditional entropy $H(X|Y):=\sum_{i=1}^k q_i \ent{\mu_i}(\rho_i)$ quantifies the volume growth of most doubly typical fibers in the following sense: 
 \begin{enumerate}
 \item   For any $\V y \in A_{\delta'}^{(n)}$, one has
 $n^{-1} \ln \mathcal H^{m(\V y)}(T^{(n)}_{\delta,\delta'}(\V y)) \leq H(X|Y) + (\delta + \delta'_n).$
 \item For any $\epsilon > 0$, the set $B_{\epsilon}^{(n)}$  of $\V y\in \subset A_{\delta'_n}^{(n)}$ such that 
 $$\frac{1}{n} \ln \mathcal H^{m(y)}(T^{(n)}_{\delta,\delta'}(\V y)) > H(X|Y) -\epsilon + (\delta + \delta'_n),$$
 satisfies 
 $$\limsup_{||(\delta,\delta'_n)||\to 0}\limsup_{n\to \infty} \frac{1}{n} \ln |B^{(n)}_\epsilon| = H(Y)= \limsup_{||(\delta,\delta'_n)||\to 0}\limsup_{n\to \infty} \frac{1}{n} \ln |A^{(n)}_{\delta_n'}|.$$
 \end{enumerate}
\end{theorem}
This gives a geometric interpretation to the possibly noninteger dimension $\mathbb E(D) = \sum_{i=1}^k q_i m_i$, which under suitable hypotheses is the information dimension of $\rho$ \cite[Sec. V]{Vigneaux-stratified}, thus answering an old question posed by Renyi in \cite[p. 209]{Renyi1959}.

\bibliographystyle{abbrv}
\bibliography{bibtex}

\end{document}